\documentclass[11pt,a4paper]{article}
\usepackage{jcappub}
\usepackage{bm}
\usepackage{amsmath}
\usepackage{mathrsfs}
\usepackage[mathscr]{euscript}
\usepackage{feynmp}
\usepackage{natbib}
\usepackage{ulem}
\usepackage{comment}

\def\Mpl{M_{\rm Pl}}

\def\0{{(0)}}
\def\sig0{\dot{\sigma}_0}

\def\ph0{\dot{\phi}_0}

\title{
Implications of the \\
Cosmic Birefringence Measurement for the Axion Dark Matter Search}
\author{Ippei Obata}
\affiliation{Max-Planck-Institut f{\"u}r Astrophysik, Karl-Schwarzschild-Str. 1, 85748 Garching, Germany}

\emailAdd{obata@mpa-garching.mpg.de}

\abstract{
We show that a recent constraint on the cosmic birefringence effect due to dark energy can be related to the constraints on the coupling of axion dark matter to photon, by relying on a simple model of two-axion alignment mechanism with periodic potentials.
Owing to the alignment of the potentials, one linear combination of two fields provides a nearly flat direction and acts as dark energy, whereas the other combination provides a steep direction and acts as dark matter.
This scenario solves the known conceptual issues of one-field model for dark energy and predicts the connection between seemingly disparate constraints on the dark sectors of our universe.}

\arxivnumber{2108.02150}

\usepackage{amsfonts}

\begin{document}

\maketitle

\section{Introduction}

Identifying the nature of dark energy and dark matter is the central issue in modern cosmology \cite{Weinbergcos}.
In this paper, we consider a possibility that these dark sectors are not independent components but are related to each other by a common origin.
This kind of idea has been developed in the framework of the unified models for dark energy and dark matter \cite{Bento:2002ps,Makler:2002jv,Carturan:2002si,Sandvik:2002jz,Scherrer:2004au,Giannakis:2005kr,Bruni:2012sn,Leanizbarrutia:2017afj,Ferreira:2018wup}.
Other than these models, pseudo scalar fields such as axions  \cite{Peccei:1977hh,Weinberg:1977ma,Wilczek:1977pj,Kim:1979if,Shifman:1979if,Dine:1981rt,Zhitnitsky:1980tq,Svrcek:2006yi,Arvanitaki:2009fg} are also well-known candidates for both of them ({\it e.g.} see \cite{Marsh:2015xka} for review).
Axions weakly couple to photons via the topological Chern-Simons coupling and violate the parity symmetry of photon's polarization states.
Hence, observing the parity-violating signals may help to understand the nature of dark energy and dark matter.

The conventional axion search has been to look for the axion-photon conversion under the background magnetic field \cite{Sikivie:1983ip,Raffelt:1987im}.
Recently, however, a new alternative method to measure the parity violation comes to receive an attention.
Since the axion differentiates the phase velocities between left- and right-handed photons via the Chern-Simons coupling,
the photon's polarization plane slightly rotates under the configuration of background axion field: namely, axion behaves as a birefringent material in our universe \cite{Carroll:1989vb,Harari:1992ea,Carroll:1998zi}.
Interestingly, the induced birefringence angle oscillates in time with a frequency of axion mass, which is quite advantageous to extract the signal from the foreground in the real-time measurements.
To test the birefringence caused by axion dark matter, several experimental methods with the use of optical resonant cavities have been proposed \cite{DeRocco:2018jwe,Obata:2018vvr,Liu:2018icu,Nagano:2019rbw,Martynov:2019azm,Nagano:2021kwx}: one of these projects has already launched a prototype experiment and is now preparing for data acquisition \cite{Michimura:2019qxr,Oshima:2021irp,Fujimoto:2021uod}.
Astrophysical polarimetric surveys can be also useful to probe the cosmic birefringence from axion dark matter such as by the observation of photoplanetary disk \cite{Fujita:2018zaj}, the Event Horizon Telescope \cite{Chen:2019fsq,Chen:2021lvo}, and the cosmic microwave background (CMB) polarization \cite{Finelli:2008jv,Lee:2013mqa,Sigl:2018fba,Fedderke:2019ajk}.
These comprehensive searches potentially cover the broad mass window of axion dark matter ranging from $10^{-22} \text{eV}$ to $10^{-8} \text{eV}$ and improve upon the current constraints on the axion-photon coupling.

Meanwhile, the above constraints on dark matter have been treated mostly separately from that on dark energy.
In this paper, we newly show a possibility that these seemingly disparate constraints on the dark sectors are related to each other through the observation of CMB cosmic birefringence \cite{Lue:1998mq,Feng:2006dp,Liu:2006uh,Lee:2014rpa,Liu:2016dcg,Lee:2016jym}.
Recently, a weak signal of the isotropic birefringence angle $\beta = 0.35 \pm 0.14 \ \text{deg} \ (1\sigma ~\text{level})$ \cite{Minami:2020odp},
which mitigates the systematic uncertainty and excludes the null hypothesis at $2.4 \sigma$ level in {\it Planck} 2018 polarization data, has been quickly gathering an attention to a tantalizing hint of axion physics \cite{Fujita:2020aqt,Fujita:2020ecn,Berghaus:2020ekh,Takahashi:2020tqv,Sangal:2021qeg,Neves:2021tbt,Nagata:2021pvc,Fung:2021wbz,Mehta:2021pwf,Nakagawa:2021nme,Jain:2021shf,Tsujikawa:2021rpg,Lague:2021frh,Reig:2021ipa,Clark:2021kze,Fung:2021fcj,Namikawa:2021gbr,Kim:2021eye,Perivolaropoulos:2021jda,Alvey:2021hjp,Choi:2021aze,Berera:2021xqa}.
As shown in \cite{Fujita:2020ecn}, one of the attractive scenarios to explain this birefringence angle is that a homogeneous axion field slowly varies in time from the last scattering surface to present.
This implies a possibility that axion is responsible for the quintessence \cite{Fukugita:1994hq,Frieman:1995pm,Kim:1998kx,Kim:1999dc,Choi:1999xn,Nomura:2000yk,Kim:2002tq,Copeland:2006wr}.
Axion can behave as the quintessence field if
its potential is nearly flat and the initial field range is greater than the Planck scale for the slow-roll condition.
However, it generically requires a trans-Planckian axion decay constant that may not be compatible with a description of effective field theory \cite{Banks:2003sx}.
Even if the decay constant is taken to be a sub-Planckian value, the initial field should start very close to the potential maximum, which usually suffers from a fine-tuning of the initial condition \cite{Choi:1999xn,Kaloper:2005aj}.

The above issues can be relaxed if the axion quintessence is realized by the contributions from the multiple axion potentials \cite{Kim:1998kx,Kim:1999dc,Kim:2002tq,Kaloper:2005aj,Kim:2009cp,Chatzistavrakidis:2012bb}
, where the alignment of the potentials makes a nearly flat direction and an effectively large field range is therefore provided.
This idea has been also explored to resolve the trans-Planckian problem in the axionic inflationary model \cite{Kim:2004rp,Dimopoulos:2005ac,Cicoli:2014sva,Czerny:2014wza,Choi:2014rja,Kappl:2014lra,Long:2014dta}.
Interestingly, other than the quintessence field, this mechanism also provides a massive field in the orthogonal steep direction that may contribute to the dark matter sector.
In this work, we explore how to confirm this scenario through the constraint on the cosmic birefringence effect.
As a first step, we perform it by employing a simple two-field model with periodic potentials \cite{Kim:2004rp}
and introduce their axion-photon couplings.
Then, we search for the discovery space of the coupling of axion dark matter to photon inferred by the cosmic birefringence measurement.

This paper is organized as follows.
In Section \hyperref[sec: setup]{2}, we present the parameter region of the single-field axion quintessence to explain the birefringence angle in \cite{Minami:2020odp}.
In Section \hyperref[two-axion]{3}, we consider an alignment model of two axions coupled to photons.
In Section \hyperref[cons]{4}, we look for the parameter region of axions to realize the measured birefringence angle and connect the constraints on the axion-photon couplings between dark energy and dark matter.
Finally, we conclude our result in Section \hyperref[conc]{5}.
Throughout this paper, we set the natural unit $\hbar=c=1$.

\section{birefringence from single-axion model} \label{sec: setup}

To begin with, we present a misalignment model of single axion field and find the parameter space to fit the measured birefringence angle reported in \cite{Minami:2020odp}.
Regarding the axion's potential, we consider a standard periodic potential generated by a quantum non-perturbative effect. It is 
typically parametrized as
\begin{equation}
V(\phi) = 
m_\phi^2f_\phi^2\left[ 1 - \cos\left(\dfrac{\phi}{f_\phi}\right) \right] \ , \label{eq: cos}
\end{equation}
where
$m_\phi$ is a mass and $f_\phi$ is a decay constant of axion.
For a small field range $\phi \ll f_\phi$, \eqref{eq: cos} reduces to the quadratic potential $V(\phi) \simeq m_\phi^2\phi^2/2$.
The homogeneous field $\phi(t)$ obeys the Klein-Gordon equation
\begin{equation}
\ddot{\phi} + 3H\dot{\phi} + m_\phi^2f_\phi\sin\left(\dfrac{\phi}{f_\phi}\right) = 0 \ , \label{eq: eomphi}
\end{equation}
where the dot represents the derivative with respect to the cosmic time $t$.
The Hubble parameter is given by the Friedmann equation
\begin{equation}
H = H_0\sqrt{\Omega_m(a^{-3}+a_{\rm eq}a^{-4}) + \Omega_\phi} \ ,
\end{equation}
where $H_0 \equiv H(t_0) \simeq 1.4\times10^{-33}\text{eV}$ is the Hubble constant at a present time $t=t_0$, $\Omega_m \simeq 0.31$ is the density parameter of matter and $a_{\rm eq} \simeq 3.0\times10^{-4}$ is the scale factor at the matter-radiation equality $t = t_{\rm eq}$ (corresponding to the redshift  $z_{\rm eq} = (1-a_{\rm eq})/a_{\rm eq} \simeq 3300$) \cite{Planck:2018vyg}.
In addition to them, we have a density parameter of axion $\Omega_\phi \equiv (\dot{\phi}^2/2 + V)/(3\Mpl^2H_0^2)$  and assume that it behaves as a quintessential dark energy.

\begin{figure}[thbp]
  \begin{minipage}{0.02\textwidth}
(a) 
\end{minipage}
\hfill
  \begin{minipage}[b]{0.45\linewidth}
    \centering
    \includegraphics[keepaspectratio, scale=0.5]{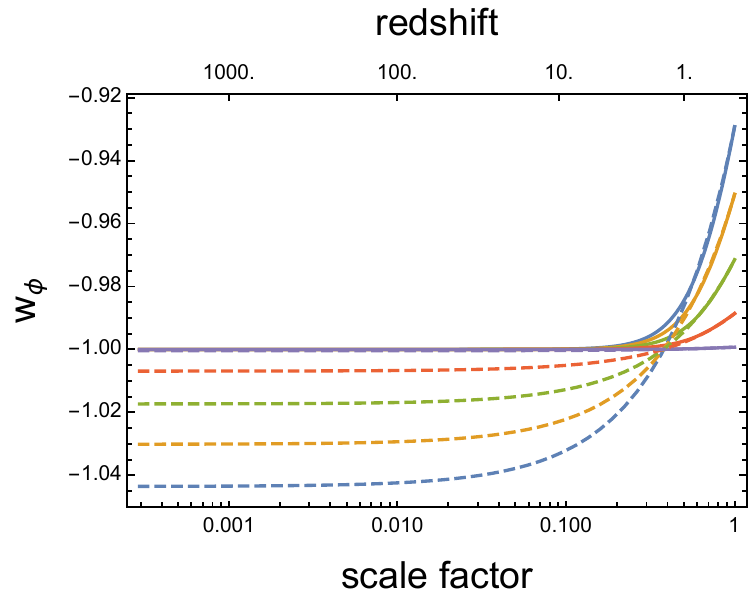}
  \end{minipage}
  \begin{minipage}{0.02\textwidth}
(b) 
\end{minipage}
 \hfill
  \begin{minipage}[b]{0.45\linewidth}
    \centering
    \includegraphics[keepaspectratio, scale=0.5]{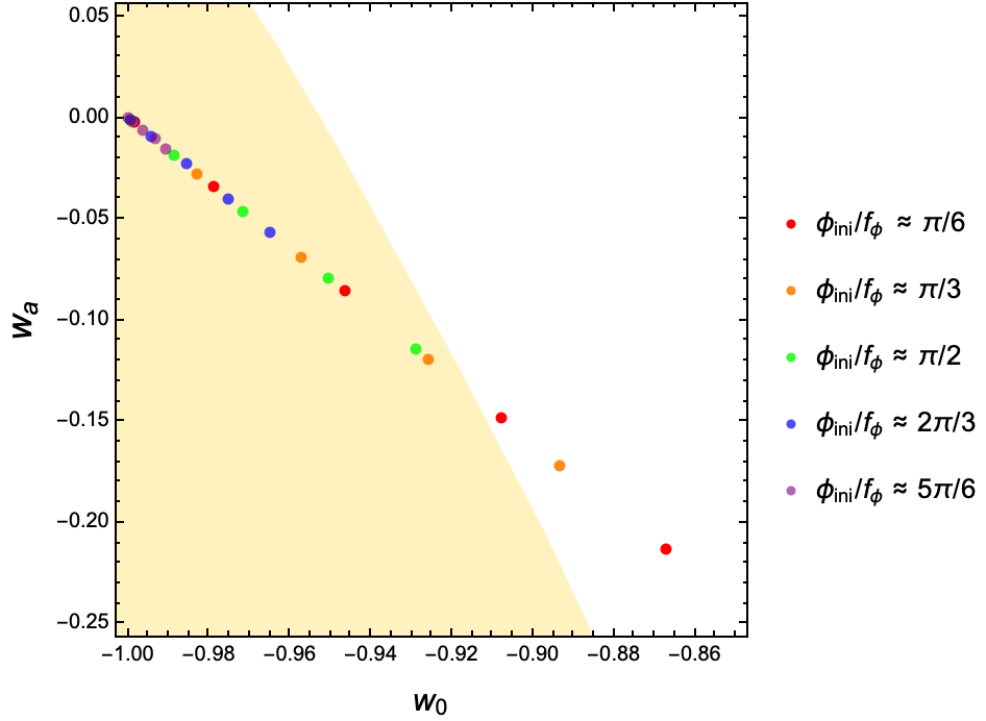}
  \end{minipage}
  \caption{(a) Time evolution of $w_\phi$ with mass $m_\phi/H_0 =1$ (blue), $m_\phi/H_0 \simeq 0.83$ (orange), $m_\phi/H_0 \simeq 0.63$ (green), $m_\phi/H_0 \simeq 0.40$ (red), and $m_\phi/H_0 = 0.1$ (purple). For the initial conditions of $\phi$ for each mass, we set the values that are very close to $\phi = \pi f_\phi/2$. The solid lines are exact solutions, while the dashed lines are the fitting solution by CPL parameterization. (b) Parameter space of $(w_0, \ w_a)$ in CPL parametrization for $\phi_{\rm ini} \simeq \pi f_\phi/6$ (red), $\phi_{\rm ini} \simeq \pi f_\phi/3$ (orange), $\phi_{\rm ini} \simeq \pi f_\phi/2$ (green), $\phi_{\rm ini} \simeq 2\pi f_\phi/3$ (blue), and $\phi_{\rm ini} \simeq 5\pi f_\phi/6$ (purple). For each initial value, we plot five dots representing the same mass parameter sets as in (a): $m_\phi/H_0 =1, \ 0.83, \ 0.63, \ 0.40, \  0.1$ (from right to left). The yellow shaded region is the $2\sigma$ constraint from $\rm{{\it Planck}} +\rm{BAO} + \rm{SNe}$ \cite{Planck:2018vyg}.}
  \label{fig:EoS}
\end{figure}

\begin{figure}[thbp]
  \begin{center}
\includegraphics[width=120mm]{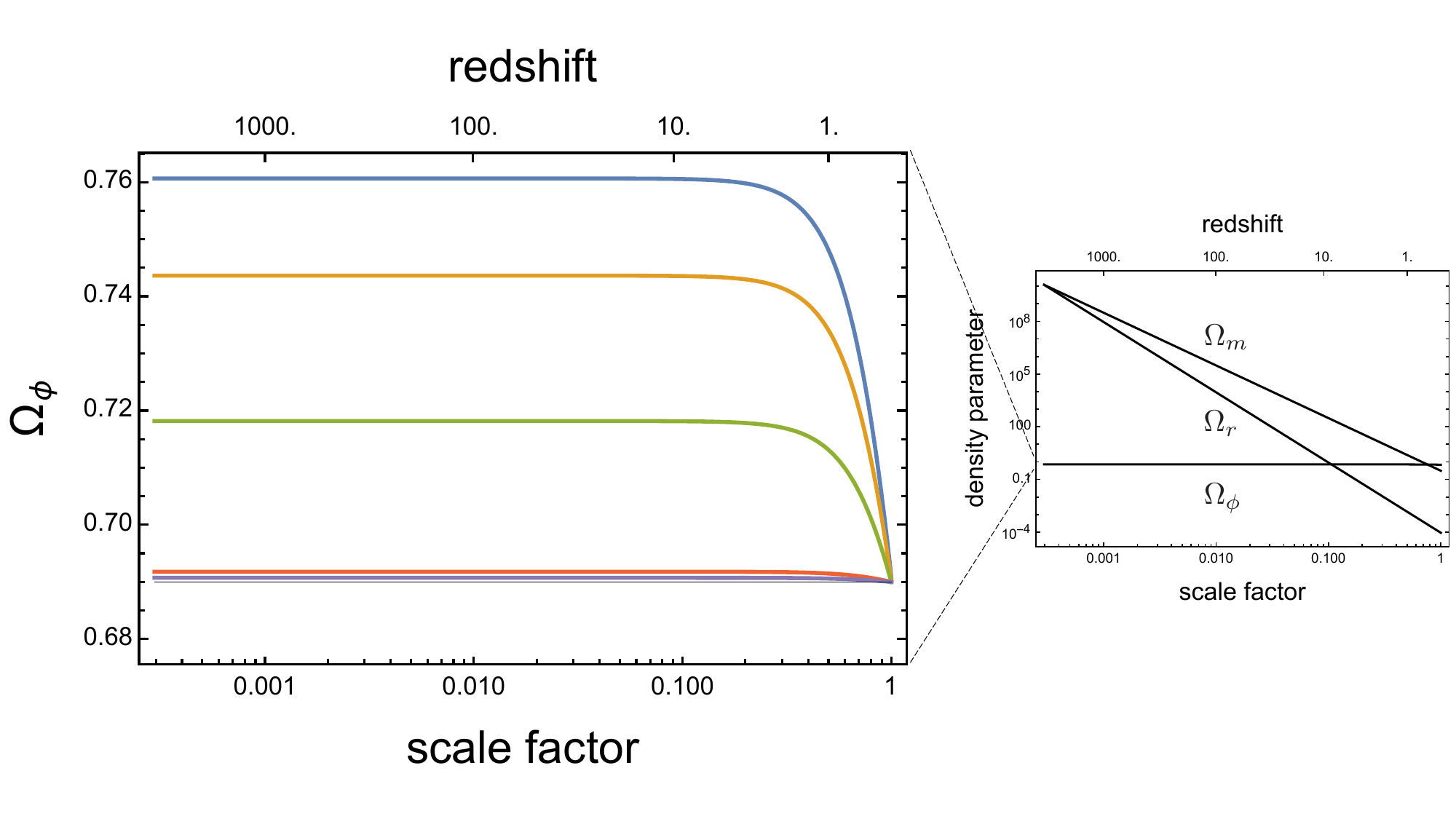}
\end{center}
 \caption{Time evolution of $\Omega_\phi$ with mass $m_\phi/H_0 =1$ (blue), $m_\phi/H_0 \simeq 0.83$ (orange), $m_\phi/H_0 \simeq 0.63$ (green), $m_\phi/H_0 \simeq 0.40$ (red), and $m_\phi/H_0 = 0.1$ (purple) for $\phi_{\rm ini} \simeq \pi f_\phi/2$. We chose the initial values and $f_\phi$ to realize $\Omega_\phi \simeq 0.69$ at present. We also show the time evolution of matter and radiation density parameters for comparison.}
  \label{fig:Omega}
\end{figure}

The time evolution of axion field is essentially determined by the ratio of the mass term to the Hubble friction term.
In the mass regime, $m_\phi \lesssim 3H_0$, the background motion is almost frozen until today because of a large Hubble friction term.
We look for the allowed parameter region of axion as dark energy by considering a mass dependence of equation of state for the axion field: $w_\phi \equiv (\dot{\phi}^2/2 - V)/(\dot{\phi}^2/2 + V)$.
Since $w_\phi$ dynamically evolves in time, we have to take care of the constraints on the dynamical dark energy, especially classified as a slow-roll thawing quintessence model \cite{Caldwell:2005tm,Scherrer:2007pu}.
To obtain a bound on axion mass, we fit the evolution of $w_\phi$ in terms of the following functional form dubbed the CPL parameterization \cite{Chevallier:2000qy,Linder:2002et}:
\begin{equation}
w_{\phi, \rm CPL} \equiv w_0 + w_a(1-a) \qquad (w_0, \ w_a: \text{const}.) \ ,
\end{equation}
which is adopted to test a time-varying equation of state in {\it Planck} mission \cite{Planck:2018vyg}.
For $w_\phi$ with several axion masses $m_\phi \lesssim H_0$, we describe its time evolution with respect to the scale factor or the redshift (left panel in Figure \ref{fig:EoS}).
In these plots, we fit $w_\phi$ by $w_{\phi, \rm CPL}$ in the redshift $0 \lesssim z \lesssim 1$ where $(w_0, w_a)$ is most strongly constrained in the whole region by Type Ia supernovae (SNe) and baryon acoustic oscillation (BAO) measurements \cite{Planck:2018vyg}.
Then, we show the parameter region of $(w_0, w_a)$ predicted by our scenario for several initial values and masses of axion field (right panel in Figure \ref{fig:EoS}).
Regarding the initial velocity of axion field, we assumed $\dot{\phi}_{\rm ini} = 0$ since the velocity is sufficiently dumped at early times.
For a large initial value, the axion field is close to the top of cosine potential and hence the slope becomes flat, leading to a tiny deviation of $(w_0, w_a)$ from the cosmological constant $(-1,0)$.
On the other hand, $(w_0, w_a)$ tends to deviate from $(-1, 0)$ for a smaller initial value of axon field.
This is because as an initial value decreases, a larger decay constant $f_\phi$ becomes necessary to produce $\Omega_\phi \simeq 0.69$, leading to a large gradient force in \eqref{eq: eomphi}.
Hereafter, to give sample solutions, we adopt the initial value $\phi_{\rm ini} \simeq \pi f_\phi/2$ consistent with the observation of dynamical dark energy.
Next, we describe the time evolution of $\Omega_\phi$ in Figure \ref{fig:Omega}.
%
The present density parameter of axion field is estimated as $\Omega_{\phi} \simeq m_\phi^2f_\phi^2/(3\Mpl^2H_0^2)$, where we have used the slow-roll approximation and neglected the kinetic energy.
For $\phi$ to be a dominant component of dark energy $\Omega_\phi \simeq 0.69$ \cite{Planck:2018vyg},
we get the following condition for the axion's decay constant
\begin{equation}
f_\phi \simeq 14\Mpl \left(\dfrac{\Omega_\phi}{0.69}\right)^{1/2}\left(\dfrac{m_\phi/H_0}{0.1}\right)^{-1} \ , \label{eq: conf}
\end{equation}
which is a trans-Planckian regime as is expected.
%
%
\begin{figure}[thbp]
  \begin{minipage}{0.02\textwidth}
(a) 
\end{minipage}
\hfill
  \begin{minipage}[b]{0.45\linewidth}
    \centering
    \includegraphics[keepaspectratio, scale=0.55]{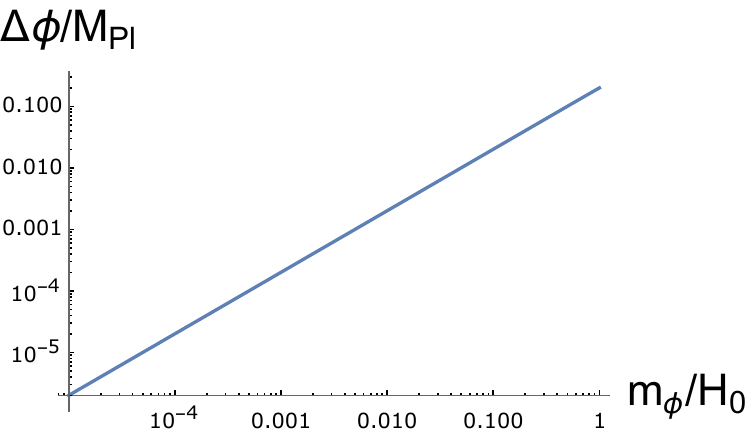}
  \end{minipage}
    \begin{minipage}{0.02\textwidth}
(b) 
\end{minipage}
\hfill
  \begin{minipage}[b]{0.45\linewidth}
    \centering
    \includegraphics[keepaspectratio, scale=0.55]{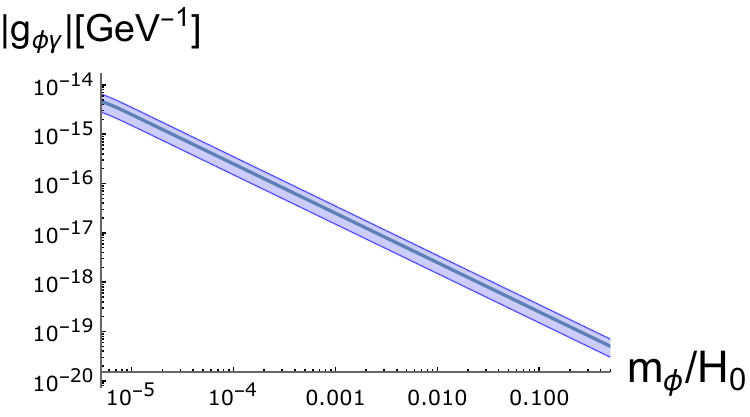}
  \end{minipage}
  \caption{(a) A numerical plot of $\Delta\phi$ with respect to axion mass $m_\phi$. (b) The resultant constraint of $g_{\phi\gamma}$ with respect to $m_\phi$ from the measurement of $\beta = 0.35\pm0.14 \ \text{deg}$ (solid blue and blue shaded region) via \eqref{eq: beta}.}
  \label{fig:deltaphigm}
\end{figure}

In the Lagrangian density, the axion-photon interaction takes the following form
\begin{equation}
\mathcal{L} \supset \dfrac{1}{4}g_{\phi\gamma}\phi F_{\mu\nu}\tilde{F}^{\mu\nu} \ ,
\end{equation}
where $\phi$ is an axion field, $g_{\phi\gamma}$ is a dimensionful coupling constant, $F_{\mu\nu}$ is a field strength of photon and $\tilde{F}^{\mu\nu}$ is its dual.
This coupling induces a rotation of polarization plane and produces a birefringence angle \cite{Harari:1992ea}
\begin{equation}
\beta = \dfrac{g_{\phi\gamma}}{2}\Delta\phi \equiv \dfrac{g_{\phi\gamma}}{2}(\phi_0 - \langle\phi_{\rm LSS}\rangle) \label{eq: beta}
\end{equation}
determined by the difference of axion field values at a present photon absorption time $\phi_0$ and a photon emission time at the last scattering surface $\phi_{\rm LSS}$ around a redshift $z \simeq 1090$.
Here, $\langle\phi_{\rm LSS}\rangle$ is an effective background value at the last scattering surface weighted by the visibility function when the CMB photons emitted \cite{Fedderke:2019ajk}.
Since the quintessence is almost frozen at this epoch, we can simply approximate $\langle\phi_{\rm LSS}\rangle \simeq \phi_{\rm LSS}$.

From \eqref{eq: beta}, we can constrain the values of $g_{\phi\gamma}(\Delta\phi)$ given that the measured birefringence angle is $\beta = 0.35\pm0.14  \ \text{deg}$ \cite{Minami:2020odp}.
At around the inflection point, the potential gradient term in \eqref{eq: eomphi} is approximately proportional to $m_\phi^2f_\phi$, so that the field excursion has a dependence $\Delta\phi \propto m_\phi^2f_\phi/H_0^2$.
Then, we obtain an approximate solution
\begin{equation}
\Delta\phi \simeq -\dfrac{1}{2}\sqrt{\dfrac{\Omega_\phi}{3}}\dfrac{m_\phi}{H_0}(H_0 t_0)^2 \label{eq: deltaap}
\end{equation}
and find that this solution agrees very well with the numerical result.
Then, we find the constraint on the axion-photon coupling $g_{\phi\gamma}$ for dark energy (see Figure \ref{fig:deltaphigm})
\begin{align}
|g_{\phi\gamma}| \simeq
2.5\times10^{-19}\text{GeV}^{-1}\left(\dfrac{\beta}{0.35\text{deg}}\right)\left(\dfrac{\Omega_\phi}{0.69}\right)^{-1/2}\left(\dfrac{m_\phi/H_0}{0.1}\right)^{-1} \ .
\label{eq: ga}
\end{align}
The amplitude of $|g_{\phi\gamma}|$ generically depends on the initial value of the axion field.
Since the gradient of the potential is steepest around the inflection point which leads to a maximum field displacement $|\Delta\phi|$, the above estimate is approximately the minimum value of $|g_{\phi\gamma}|$. 

Finally, we mention an implication of the above constraint.
The axion-photon coupling constant is represented by
\begin{equation}
g_{\phi\gamma} =\dfrac{\alpha}{2\pi}\dfrac{c_{\phi\gamma}}{f_\phi} \ . \label{eq: ca}
\end{equation}
Here, $\alpha$ is a fine-structure constant given by $\alpha = e^2/(4\pi) \simeq 1/137$.
The dimensionless constant $c_{\phi\gamma}$ is usually given by the ratio of anomaly coefficients of gauge groups and hence its value is model-dependent.
In the standard quantum chromodynamics (QCD) axion models, it takes order unity: $c_{\phi\gamma, \rm KSVZ} \sim -2$ (KSVZ model \cite{Kim:1979if,Shifman:1979if}) and $c_{\phi\gamma, \rm DFSZ} \sim 0.7$ (DFSZ model \cite{Dine:1981rt,Zhitnitsky:1980tq}).
Plugging \eqref{eq: conf} and \eqref{eq: ga} into \eqref{eq: ca}, we obtain the expected amplitude of $c_{\phi\gamma}$ from the birefringence measurement
\begin{align}
|c_{\phi\gamma}| \simeq 
7.5\times10^3\left(\dfrac{\beta}{0.35\text{deg}}\right)\left(\dfrac{m_\phi/H_0}{0.1}\right)^{-2}
 \ . \label{eq: N}
\end{align}
Therefore, the birefringence constraint requires a huge coupling constant for dark energy.

\section{Two-axion model with photon couplings}
\label{two-axion}

Next, we consider a model of two-axion fields $(\phi, \chi)$ with cosine potentials to relax the conceptual issues mentioned in the previous section.
The potential form is as follows \cite{Kim:2004rp}:
\begin{align}
V(\phi, \chi) &= \Lambda_1^4\left[1-\cos\left(\dfrac{\phi}{F_{\phi1}} + \dfrac{\chi}{F_{\chi1}} \right)\right] + \Lambda_2^4\left[1-\cos\left(\dfrac{\phi}{F_{\phi2}} + \dfrac{\chi}{F_{\chi2}} \right)\right] \ . \label{eq: KNP}
\end{align}
Regarding the decay constants $\{ F_{\phi1}, \ F_{\phi2}, \ F_{\chi1}, \ F_{\chi2} \}$, we require that they are all within the Planck scale $\Mpl$.
We can see that when the following condition
\begin{equation}
\dfrac{F_{\chi1}}{F_{\phi1}} = \dfrac{F_{\chi2}}{F_{\phi2}} \label{eq: flat}
\end{equation}
holds, the same field combination appears in both cosine potentials in \eqref{eq: KNP}: namely, an exactly-flat direction appears in the potential.
Then, a nearly flat potential is realized if the relationship \eqref{eq: flat} is slightly broken by a small parameter.
We define it as
\begin{equation}
F_{\chi2} = F_{\chi1}(1+\epsilon)
\end{equation}
and assume that $\epsilon$ is much smaller than unity: $\epsilon \ll 1$.

\begin{figure}[thbp]
  \begin{minipage}{0.02\textwidth}
(a) 
\end{minipage}
\hfill
  \begin{minipage}[b]{0.45\linewidth}
    \centering
    \includegraphics[keepaspectratio, scale=0.25]{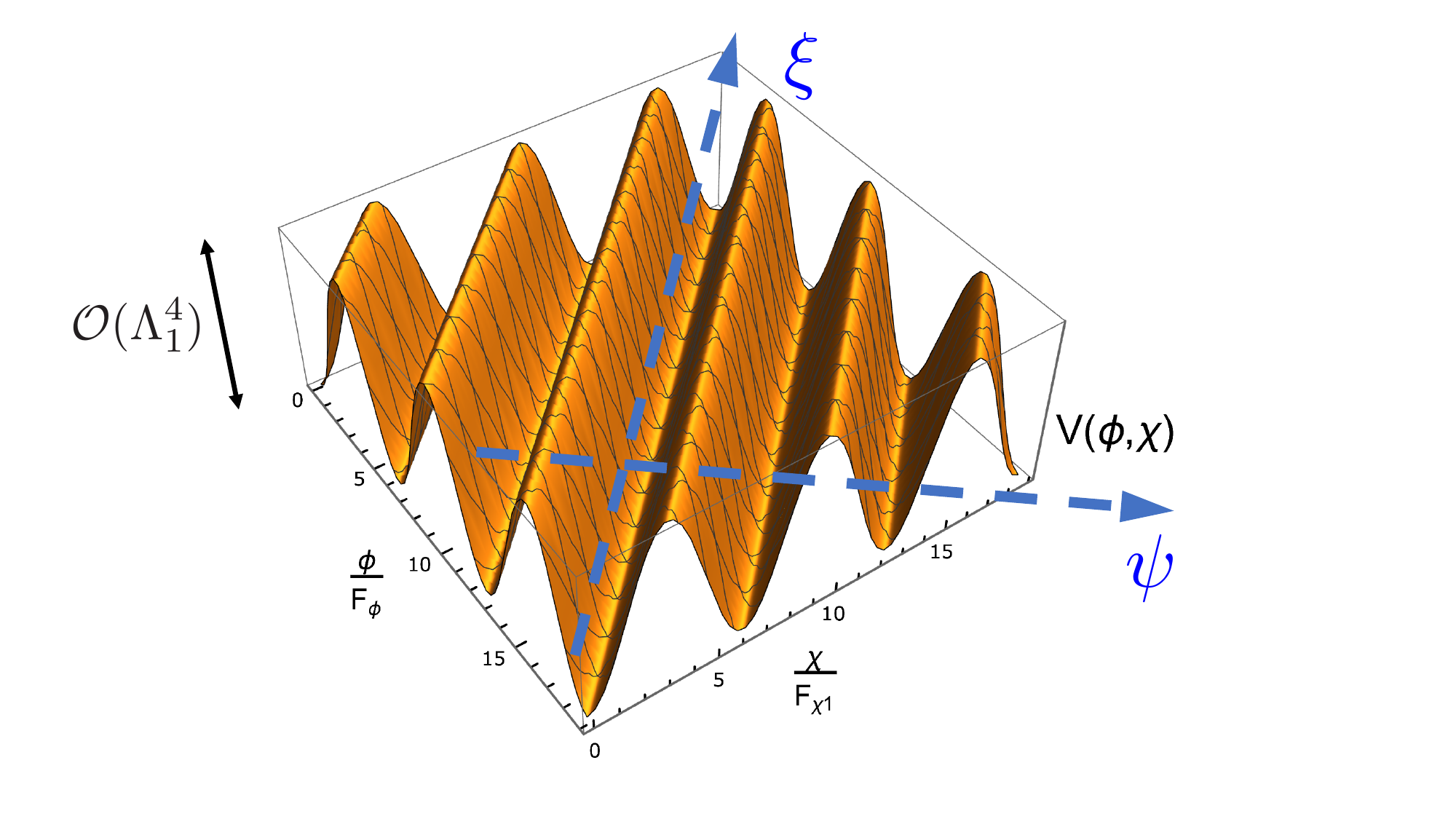}
  \end{minipage}
    \begin{minipage}{0.02\textwidth}
(b) 
\end{minipage}
\hfill
  \begin{minipage}[b]{0.45\linewidth}
    \centering
    \includegraphics[keepaspectratio, scale=0.22]{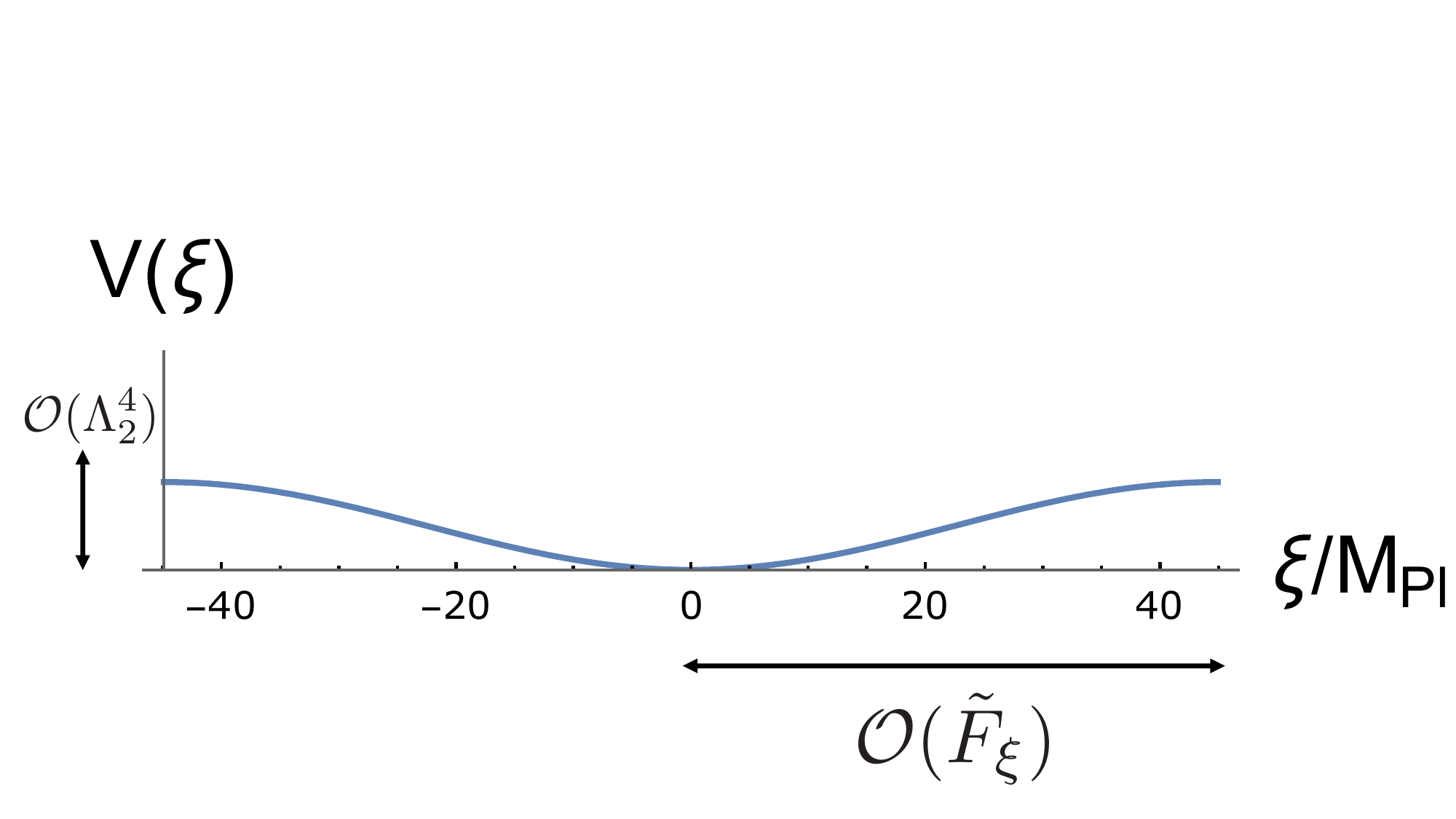}
  \end{minipage}
  \caption{(a) The superposition of periodic potentials \eqref{eq: KNP} provides a nearly-massless field $\xi$ and a massive field $\psi$. (b) The effective potential of $\xi$. In this plot, we set $\tilde{F}_\xi \simeq 14\Mpl.$}
  \label{fig:potential}
\end{figure}

To find the mass eigenstate of this system, let us transform the basis $(\phi, \ \chi)$. 
We expand the potential near the origin of axon fields $(\phi =\chi = 0)$ and express it in terms of a Hessian matrix
\begin{align}
V(\phi, \chi)
&\simeq \dfrac{1}{2}(\begin{array}{cc}
\phi & \chi
\end{array})\left(\begin{array}{cc}
V_{\phi\phi} & V_{\phi\chi} \\
V_{\chi\phi} & V_{\chi\chi}
\end{array}\right)\left(\begin{array}{c}
\phi \\
\chi
\end{array}\right)
 \ ,
\end{align}
where the indices $\{ \phi, \chi \}$ of the potential are the field derivatives.
For our analytical convenience, hereafter we set $F_{\phi1} = F_{\phi2} \equiv F_\phi$ and assume a hierarchy of the potential scales $\Lambda_1^4 \gg \Lambda_2^4$.
Then, at the leading order of $\epsilon$, we can find the following two mass eigenvalues
\begin{align}
m^2_\xi &\simeq \dfrac{\Lambda_2^4}{F_\phi^2+F_{\chi1}^2}\epsilon^2 \ , \qquad
m_\psi^2 \simeq \dfrac{F_\phi^2+F_{\chi1}^2}{F_\phi^2F_{\chi1}^2}\Lambda_1^4 
\end{align}
and the associated new coordinates $(\xi, \ \psi)$:
\begin{align}
\xi &\simeq \dfrac{F_\phi}{\sqrt{F_\phi^2+F_{\chi1}^2}}\phi - \dfrac{F_{\chi1}}{\sqrt{F_\phi^2+F_{\chi1}^2}}\chi \ , \label{eq: trans1} \\
\psi &\simeq -\dfrac{F_{\chi1}}{\sqrt{F_\phi^2+F_{\chi1}^2}}\phi -\dfrac{F_\phi}{\sqrt{F_\phi^2+F_{\chi1}^2}}\chi \ , \label{eq: trans2}
\end{align}
which configure the eigenbasis of mass matrix.
In terms of $\psi$ and $\xi$, the total potential is approximately rewritten as
\begin{align}
V(\psi, \ \xi) \simeq &\Lambda_1^4\left[ 1 - \cos\left(\dfrac{\sqrt{F_\phi^2+F_{\chi1}^2}}{F_\phi F_{\chi1}}\psi\right) \right] \notag \\
+&\Lambda_2^4\left[ 1 - \cos\left(\dfrac{\sqrt{F_\phi^2+F_{\chi1}^2}}{F_\phi F_{\chi1}}\psi - \dfrac{\epsilon}{\sqrt{F_\phi^2+F_{\chi1}^2}}\xi \right) \right] \ .
\end{align}
From this potential, we obtain their effective field ranges
\begin{equation}
\tilde{F}_{\xi} \simeq \dfrac{\sqrt{F_\phi^2+F_{\chi1}^2}}{\epsilon} \ , \qquad \tilde{F}_\psi \simeq \dfrac{F_\phi F_{\chi1}}{{\sqrt{F_\phi^2+F_{\chi1}^2}}} \ . \label{eq: efr}
\end{equation}
For $\tilde{F}_\xi$, it becomes greater and exceeds a Planck scale as $\epsilon$ goes to zero even if $F_{\phi, \ \chi1} \leq \Mpl$.
On the other hand, $\tilde{F}_\psi$ is a sub-Planckian value as long as $F_{\phi, \ \chi1} \leq \Mpl$.
Their mass ratio is evaluated as
\begin{equation}
\dfrac{m_\psi}{m_\xi} \simeq \dfrac{1}{\epsilon}
\dfrac{\Lambda_1^2}{\Lambda_2^2}\dfrac{F_\phi^2+F_{\chi1}^2}{F_\phi F_{\chi1}}  \gg 1 \ .
\end{equation}
Therefore, $\psi$ always possesses a heavier mass and a smaller decay constant in comparison with $\xi$.
In Figure \ref{fig:potential}, we illustrate the aligned potential of $\xi$ and $\psi$ and describe the nearly flat potential of $\xi$.

Finally, to relate the above alignment model with the cosmic birefringence, we introduce the couplings of photons to the original axion fields
\begin{equation}
\mathcal{L} \supset \dfrac{\alpha}{8\pi}\left(\dfrac{\phi}{F_{\phi\gamma}} + \dfrac{\chi}{F_{\chi\gamma}} \right)F_{\mu\nu}\tilde{F}^{\mu\nu} \ . \label{eq: axipho}
\end{equation}
Notice that we have absorbed the dimensionless coefficients (which corresponds to $c_{\phi\gamma}$ in \eqref{eq: ca}) into the definition of decay constants $F_{\phi\gamma}, \ F_{\chi\gamma}$. 
In terms of \eqref{eq: trans1} and \eqref{eq: trans2},
we can rewrite \eqref{eq: axipho} as the axion-photon couplings for $\xi$ and $\psi$.
Their coupling constants are found to be
\begin{align}
 g_{\xi\gamma} &= \dfrac{\alpha}{2\pi}\dfrac{c_{\xi\gamma}}{\tilde{F}_\xi}\ , \quad c_{\xi\gamma} \equiv 
\dfrac{1}{\epsilon}\left(\dfrac{F_\phi}{F_{\phi\gamma}} - \dfrac{F_{\chi1}}{F_{\chi\gamma}}\right)  \ , \label{eq: cxi} \\
g_{\psi\gamma} &= \dfrac{\alpha}{2\pi}\dfrac{c_{\psi\gamma}}{\tilde{F}_\psi} \ , \quad c_{\psi\gamma} \equiv
 - \left(\dfrac{F_\phi}{F_{\chi\gamma}} + \dfrac{F_{\chi1}}{F_{\phi\gamma}}\right)\dfrac{F_\phi F_{\chi1}}{F_\phi^2+F_{\chi1}^2} \label{eq: cpsi} \ .
\end{align}
The dependence of the coefficient $c_{\xi\gamma}$ on $\epsilon$ arises from the normalization of the effective field range $\tilde{F}_{\xi} \propto \epsilon^{-1}$.
Hence, $c_{\xi\gamma}$ is also enhanced as we take a small $\epsilon$.
In addition, these coefficients depend on the ratios of decay constants.
To make things clear, defining the following parameters
\begin{equation}
a \equiv F_{\chi1}/F_\phi \ , \quad b \equiv F_{\phi}/F_{\phi\gamma} \ , \quad c \equiv F_{\chi1}/F_{\chi\gamma} \ ,
\end{equation}
\eqref{eq: cxi} and \eqref{eq: cpsi} are represented by
\begin{equation}
c_{\xi\gamma} = \dfrac{b-c}{\epsilon} \ , \qquad c_{\psi\gamma} = -\dfrac{a^2b+c}{1+a^2} \ . \label{eq: c}
\end{equation}
Below this section, we constrain the parameter regions of $g_{\xi\gamma}$ and $g_{\psi\gamma}$ as dark energy and dark matter from the measurement of cosmic birefringence. 
%

\section{Parameter Search}
\label{cons}

We explore the parameter space of two axion fields consistent with the observational conditions and connect the constraints on the photon couplings between dark energy and dark matter.
Firstly, we focus on the parameter region of $\xi$ as the quintessence producing the cosmic birefringence angle.
The condition for the effective field range $\tilde{F}_\xi$ is obtained by replacing the parameters of $\phi$ in \eqref{eq: conf} with that of $\xi$.
In terms of the definition of $\tilde{F}_\xi$ \eqref{eq: efr},
we obtain the condition for $\epsilon$:
\begin{align}
\epsilon &\simeq 7.0\times10^{-2}\dfrac{\sqrt{F_\phi^2 + F_{\chi1}^2}}{\Mpl}
\left(\dfrac{\Omega_\xi}{0.69}\right)^{-1/2}\left(\dfrac{m_\xi/H_0}{0.1}\right) \ . \label{eq: eps}
\end{align}
We notice that $\sqrt{F_\phi^2+F_{\chi1}^2}/\Mpl \lesssim \sqrt{2}$ holds from the requirement of sub-Planckian values $F_{\phi, \ \chi1} \leq \Mpl$.
Consequently, $\epsilon$ should be less than $\mathcal{O}(10^{-1})$ for $\xi$ to be quintessence.
We also get the constraint on $c_{\xi\gamma}$ from the birefringence angle by replacing $c_{\phi\gamma}$ in \eqref{eq: N} with $c_{\xi\gamma}$.
Since $c_{\xi\gamma} = (b-c)/\epsilon$, plugging \eqref{eq: eps} into \eqref{eq: cxi}, we obtain the constraint on the ratios of decay constants
\begin{align}
|b-c| &\simeq 5.2\times10^2\dfrac{\sqrt{F_\phi^2 + F_{\chi1}^2}}{\Mpl}\left(\dfrac{\beta}{0.35\text{deg}}\right)\left(\dfrac{\Omega_\xi}{0.69}\right)^{-1/2}\left(\dfrac{m_\xi/H_0}{0.1}\right)^{-1} \ , \label{eq: fphi}
\end{align}
which leads to the constraint on $c_{\psi\gamma}$ via \eqref{eq: c}.

For $\psi$, we recognize it as the dark matter and consider the mass range $m_\psi \gtrsim 10^{-22}\text{eV}$ including the region of fuzzy dark matter \cite{Hui:2016ltb}.
We assume a simple misalignment production mechanism of axion dark matter \cite{Preskill:1982cy,Abbott:1982af,Dine:1982ah,Arias:2012az}, thereby obtaining the abundance of $\psi$ as \cite{Marsh:2010wq}
\begin{align}
\Omega_\psi \simeq \dfrac{1}{6}\left(\dfrac{\tilde{F}_\psi}{\Mpl}\right)^2(9\Omega_r)^{3/4}\left(\dfrac{m_\psi}{H_0}\right)^{1/2} \ ,
\end{align}
where $\Omega_r$ is the density parameter of radiation.
From this equation, we get the condition for $\tilde{F}_\psi$:
\begin{equation}
\tilde{F}_\psi \simeq 3.8\times10^{-2}\Mpl\left(\dfrac{\Omega_\psi}{0.31}\right)^{1/2}\left(\dfrac{m_\psi}{10^{-22}\text{eV}}\right)^{-1/4} \ . \label{eq: DM}
\end{equation}

We are interested in the parameter space to predict a large value of $g_{\psi\gamma}$ consistent with the birefringence constraint \eqref{eq: fphi}.
For our convenience, we consider the case $c=0$ where $\chi$ does not couple to the photon and divide the parameter space into the three regions with respect to $a$.

\subsection{region (i): $a = \mathcal{O}(1)$}

We firstly consider the region where $a = F_{\chi1}/F_\phi$ is around an order unity.
At this time, the coefficient $c_{\psi\gamma}$ is given by $c_{\psi\gamma}\sim -b/2$ from the relationship \eqref{eq: c}.
Since $F_\phi$ and $F_{\chi1}$ are comparable, we can approximate $\sqrt{F_\phi^2 + F_{\chi1}^2} \sim 2\tilde{F}_\psi$ in \eqref{eq: fphi}.
Then, $g_{\psi\gamma}$ is roughly evaluated as
\begin{align}
|g_{\psi\gamma}| = \dfrac{\alpha}{2\pi}\dfrac{|c_{\psi\gamma}|}{\tilde{F}_\psi} &\simeq 2.5\times10^{-19} ~\text{GeV}^{-1}\left(\dfrac{\beta}{0.35\text{deg}}\right)\left(\dfrac{\Omega_\xi}{0.69}\right)^{-1/2}\left(\dfrac{m_\xi/H_0}{0.1}\right)^{-1} \ .
\end{align}
We notice that the dependence of $g_{\psi\gamma}$ on $\tilde{F}_\psi$ vanishes and the resultant constraint is independent on the mass of axion dark matter.

\subsection{region (ii): $a \gg 1$}

Next, we consider the parameter region of $F_{\phi} \ll F_{\chi1}$,
where $F_\psi$ and $c_{\psi\gamma}$ are reduced to
$\tilde{F}_\psi \simeq F_{\phi}, \ c_{\psi\gamma} \simeq -b$.
Then, from \eqref{eq: fphi} and \eqref{eq: DM}, we obtain
\begin{align}
|g_{\psi\gamma}| &\simeq 6.6 \times 10^{-18} ~\text{GeV}^{-1}\dfrac{F_{\chi1}}{\Mpl}\left(\dfrac{\Omega_\xi}{0.69}\right)^{-1/2}\left(\dfrac{\Omega_\psi}{0.31}\right)^{-1/2} \notag \\
&\times\left(\dfrac{\beta}{0.35\text{deg}}\right)\left(\dfrac{m_\xi/H_0}{0.1}\right)^{-1}\left(\dfrac{m_\psi}{10^{-22}\text{eV}}\right)^{1/4} \ . \label{eq: main}
\end{align}
For $F_{\chi1}$, it takes a much greater value than $\tilde{F}_\psi$ \eqref{eq: DM} from the condition $a \gg 1$.
Therefore, this region predicts a greater amplitude of $g_{\psi\gamma}$ in comparison with the region (i) and $|g_{\psi\gamma}|$ increases for a heavier mass range of axion dark matter.

\subsection{region (iii): $a \ll 1$}

Finally, we consider the parameter region of $F_{\phi} \gg F_{\chi1}$,
where $\tilde{F}_\psi$ and $c_{\psi\gamma}$ are reduced to
$\tilde{F}_\psi \simeq F_{\chi1}, \ c_{\psi\gamma} \simeq -a^2b$.
In comparison with the region (ii), the amplitude is always suppressed by a factor of $a^2 \ll 1$.

\vskip\baselineskip

\begin{figure}[t]
\begin{center}
\includegraphics[width=130mm]{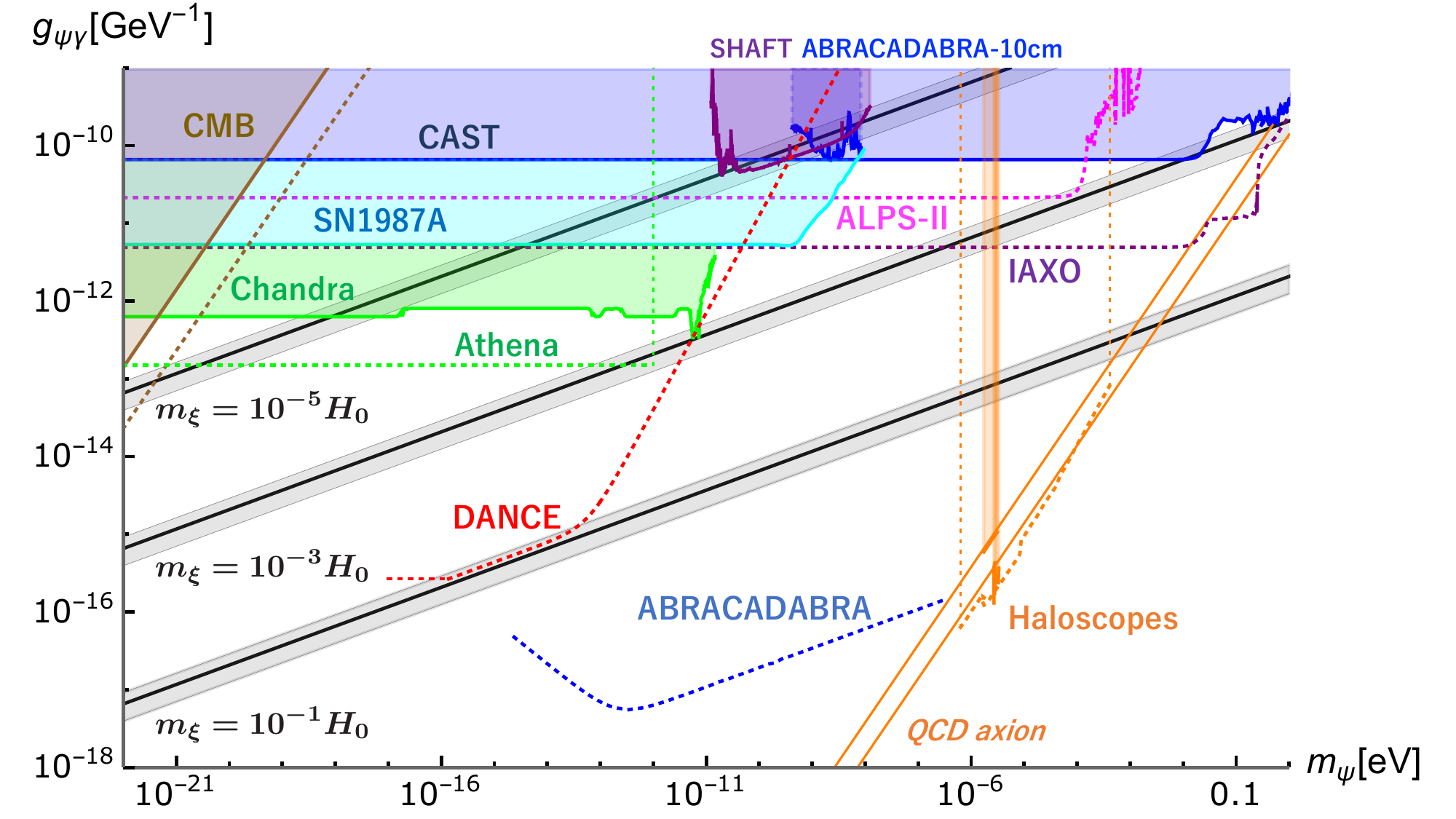}
\end{center}
\caption{
Parameter space for the axion-photon coupling $g_{\psi\gamma}$ inferred by the measurement of birefringence angle $\beta = 0.35\pm0.14 \ \text{deg}$ versus the axion mass $m_\psi$.
The solid black lines with gray bands are the axion dark matter regions (ii) with several values of axion quintessence mass $m_\xi$: from $10^{-1}H_0$ (bottom) to $10^{-5}H_0$ (top).
We choose $F_{\chi1} = \Mpl$ and assume that $\xi$ and $\psi$ are the dominant components of dark energy $\Omega_\xi = 0.69$ and dark matter $\Omega_\psi = 0.31$.
The orange lines are the parameter region of QCD axion models \cite{Kim:1979if,Shifman:1979if,Dine:1981rt,Zhitnitsky:1980tq}.
The other shaded regions are excluded by the measurements of CAST \cite{Zioutas:2004hi,Anastassopoulos:2017ftl} (blue), ADMX \cite{ADMX:2009iij,ADMX:2018gho,ADMX:2020hay} (orange), SHAFT \cite{Gramolin:2020ict} (purple), ABRACADABRA-10cm \cite{Salemi:2021gck} (dark blue), SN1987A \cite{Payez:2014xsa} (cyan), Chandra \cite{Reynolds:2019uqt} (green), and {\it Planck} \cite{Fedderke:2019ajk} (brown).
The dotted lines show the areas within the reach of future measurements of IAXO \cite{Vogel:2013bta} (purple), QCD axion haloscopes \cite{Irastorza:2018dyq} (orange), ALPS-II \cite{Bahre:2013ywa} (magenta), ABRACADABRA \cite{Kahn:2016aff} (blue), DANCE \cite{Obata:2018vvr,Michimura:2019qxr} (red), Athena \cite{Conlon:2017ofb} (green), and CMB polarizations \cite{Fedderke:2019ajk} (brown).
}
\label{fig:gpsi}
\end{figure}

To conclude, we find that $g_{\psi\gamma}$ becomes largest when there is a hierarchy of decay constants $F_\phi \ll F_{\chi1}$ (region (ii)).
In Figure \ref{fig:gpsi}, we illustrate the expected parameter regions of $g_{\psi\gamma}$ with respect to $m_\psi$ ranging from $10^{-22} ~\text{eV}$ to $1~\text{eV}$ as a window of ultralight dark matter \cite{Ferreira:2020fam}.
We can see that, for a smaller mass of quintessence $m_\xi$ (corresponding to a smaller value of $\epsilon$), a greater coupling $g_{\psi\gamma}$ is predicted.
With several scales of $m_\psi$, we can test these parameter regions by a variety of measurements for the axion searches.

Finally, we shortly comment on the parameter region $c \neq 0$.
When $b = 0$, we can obtain the same results as the above cases just by replacing $a$ with $1/a$: namely, the largest coupling arises from the branch $F_{\chi1} \ll F_\phi$.
In the case of $b \neq 0$ and $c \neq 0$, the predictions are a bit complicated but similar, except for the case when $b$ and $c$ take the almost same values.
Since $c_{\psi\gamma}$ \eqref{eq: c} is given by an order of $b$ or $c$, $g_{\psi\gamma}$ potentially gets larger than \eqref{eq: main} if a cancellation happens in $|b-c|$ \eqref{eq: fphi} with many orders of magnitude.
However, it tends to drive a fine-tuning of model parameters and therefore we don't take this possibility so serious.

%
%

\subsection{Discussion}

At the end of this section, we discuss a couple of possibilities that we need to check and ensure our proposal.

\subsubsection*{QCD axion}

Intriguingly, the resultant parameter region of heavier axion field $\psi$ overlaps the parameter space of QCD axion.
In particular, we have found that our scenario predicts the typical mass window of QCD axion dark matter $10^{-6} \ \text{eV} \lesssim m_\psi \lesssim 10^{-3} \ \text{eV}$ (corresponding to $1.4\times10^{12} \ \text{GeV} \lesssim \tilde{F}_\psi \lesssim 7.6\times10^{12} \ \text{GeV}$) for $F_{\chi1} = 10^{-1.5} \ \Mpl$ and $10^{-2} \lesssim m_\xi/H_0 \lesssim 1$ (corresponding to $140 \Mpl \gtrsim \tilde{F}_\xi \gtrsim 1.4 \Mpl$ and $160 \gtrsim c_{\psi\gamma} \gtrsim 1.6)$.
It would be interesting to develop the possibility that the heavier axion field $\psi$ could realize the QCD axion.
In our current scenario, however, we have assumed that $\psi$ is generated in a manner of axion-like particle and hence does not couple to QCD sector.
For instance, to identify $\psi$ with QCD axion dark matter, introducing the finite-temperature effect of cosine potential would be necessary \cite{Gross:1980br,Wantz:2009it}.
Once we introduce this effect, the relationship between $m_\psi$ and $\tilde{F}_\psi$ changes and would modify the overlapping parameter region of QCD axion dark matter.
Then, it would be also necessary to think of how sensitive the resultant $\psi$ could influence the original QCD scenario, such as modifying the potential minimum and spoiling the solution of strong CP problem by a contribution from additional phase factor or by an interaction of dark energy to QCD axion.
We leave further consideration of these issues to future study.

\subsubsection*{Isocurvature mode}

We also mention the contribution of both fields to their isocurvature modes.
The isocurvature perturbation is given by the ratio of the amplitude of quantum fluctuation, represented by the inflationary Hubble scale $H_I/(2\pi)$, to the field range of axions.
In the case of dark energy ($\xi$ field), the mean field is super-Planckian $\xi \sim \tilde{F}_\xi \gtrsim \Mpl$ and therefore its contribution is negligible.
On the other hand, the CDM isocurvature density is tightly constrained by the CMB observations.
The amplitude of isocurvature power spectrum $A_i$ is evaluated as \cite{Marsh:2015xka}
\begin{equation}
A_i = \left(\dfrac{\Omega_\psi}{\Omega_m}\right)^2\dfrac{H_I^2}{\pi^2\psi^2} \label{eq: iso}
\end{equation}
and current {\it Planck} measurement tells us $A_i \lesssim 0.8 \times 10^{-10}$ \cite{Planck:2018jri}.
Since $\psi \sim \tilde{F}_\psi$ given by \eqref{eq: DM}, from \eqref{eq: iso} we obtain the upper bound on the inflationary Hubble scale:
\begin{equation}
H_I \lesssim 2.6\times10^{12} \ \text{GeV}\left(\dfrac{m_\psi}{10^{-22} \ \text{eV}}\right)^{-1/4} \ .
\end{equation}

\section{Conclusion}
\label{conc}

In this work, we explored the implication of CMB cosmic birefringence measurement due to dark energy and predicted the connection between the constraints on the axion-photon coupling of dark energy and dark matter.
More precisely, we considered a two-axion alignment model and introduced their photon couplings to realize the measured birefringence angle.
Owing to the alignment of the potentials, this scenario predicts two combinations of axion fields: one is an almost massless field along a nearly flat direction of the potential, and the other is a massive field along the orthogonal steep direction.
The former can play a role of quintessence field.
Accordingly, it provides a large coupling of axion quintessence to photon controlled by a small parameter of the potential misalignment.
On the other hand, the latter heavier field can contribute to the dark matter.
Depending on the smallness of the mass of quintessence field, axion dark matter can possess a sizable amount of photon coupling via the constraint on the birefringence effect for dark energy, which is potentially testable with the current or future axion dark matter observations.

To realize the measured birefringence angle for the lower quintessence mass range, the larger hierarchy of decay constants $F_\phi \gg F_{\phi\gamma}$ was required (see \eqref{eq: fphi}).
This condition generically leads to a large anomaly coefficient of axion-photon coupling: $|b| \gg 1$ or $|c| \gg 1$, which induces the matter fields with large charges and the theory may be out of the perturbative regime \cite{Agrawal:2018mkd}.
We expect that this issue could be mitigated by extending the model to include other mechanisms of enhancing the couplings such as the kinetic mixing of multiple axion fields \cite{Agrawal:2018mkd,Babu:1994id,Bachlechner:2014hsa,Shiu:2015uva}.
It would be also worth developing the compatibility of axion quintessence models with the swampland criteria in string theory \cite{Blumenhagen:2018hsh,Cicoli:2018kdo,Ibe:2018ffn,Palti:2019pca,Banerjee:2020xcn}.
We leave these issues in our future work.

\begin{acknowledgments}

We would like to thank Eiichiro Komatsu and Elisa Ferreira for the fruitful discussion and helpful comments.
This work is supported by JSPS Overseas Research Fellowship and JSPS KAKENHI Grant Number JP20H05859.

\end{acknowledgments}


\begin{thebibliography}{99}



\bibitem{Weinbergcos}
S. Weinberg, {\it Cosmology} (Oxford University Press, 2008).


\bibitem{Bento:2002ps}
M.~C.~Bento, O.~Bertolami and A.~A.~Sen,
Phys. Rev. D \textbf{66}, 043507 (2002)
[arXiv:gr-qc/0202064 [gr-qc]].

\bibitem{Makler:2002jv}
M.~Makler, S.~Quinet de Oliveira and I.~Waga,
Phys. Lett. B \textbf{555}, 1 (2003)
[arXiv:astro-ph/0209486 [astro-ph]].

\bibitem{Carturan:2002si}
D.~Carturan and F.~Finelli,
Phys. Rev. D \textbf{68}, 103501 (2003)
[arXiv:astro-ph/0211626 [astro-ph]].

\bibitem{Sandvik:2002jz}
H.~Sandvik, M.~Tegmark, M.~Zaldarriaga and I.~Waga,
Phys. Rev. D \textbf{69}, 123524 (2004)
[arXiv:astro-ph/0212114 [astro-ph]].


\bibitem{Scherrer:2004au}
R.~J.~Scherrer,
Phys. Rev. Lett. \textbf{93}, 011301 (2004)
[arXiv:astro-ph/0402316 [astro-ph]].

\bibitem{Giannakis:2005kr}
D.~Giannakis and W.~Hu,
Phys. Rev. D \textbf{72}, 063502 (2005)
[arXiv:astro-ph/0501423 [astro-ph]].


\bibitem{Bruni:2012sn}
M.~Bruni, R.~Lazkoz and A.~Rozas-Fernandez,
Mon. Not. Roy. Astron. Soc. \textbf{431}, 2907-2916 (2013)
[arXiv:1210.1880 [astro-ph.CO]].

\bibitem{Leanizbarrutia:2017afj}
I.~Leanizbarrutia, A.~Rozas-Fern\'andez and I.~Tereno,
Phys. Rev. D \textbf{96}, no.2, 023503 (2017)
[arXiv:1706.01706 [astro-ph.CO]].


\bibitem{Ferreira:2018wup}
E.~G.~M.~Ferreira, G.~Franzmann, J.~Khoury and R.~Brandenberger,
JCAP \textbf{08}, 027 (2019)
[arXiv:1810.09474 [astro-ph.CO]].


\bibitem{Peccei:1977hh}
R.~D.~Peccei and H.~R.~Quinn,
Phys. Rev. Lett. \textbf{38}, 1440-1443 (1977).

\bibitem{Weinberg:1977ma}
S.~Weinberg,
Phys. Rev. Lett. \textbf{40}, 223-226 (1978).

\bibitem{Wilczek:1977pj}
F.~Wilczek,
Phys. Rev. Lett. \textbf{40}, 279-282 (1978).


\bibitem{Kim:1979if}
J.~E.~Kim,
Phys. Rev. Lett. \textbf{43}, 103 (1979).

\bibitem{Shifman:1979if}
M.~A.~Shifman, A.~I.~Vainshtein and V.~I.~Zakharov,
Nucl. Phys. B \textbf{166}, 493-506 (1980)


\bibitem{Dine:1981rt}
M.~Dine, W.~Fischler and M.~Srednicki,
Phys. Lett. B \textbf{104}, 199-202 (1981).

\bibitem{Zhitnitsky:1980tq}
A.~R.~Zhitnitsky,
Sov. J. Nucl. Phys. \textbf{31}, 260 (1980).



\bibitem{Svrcek:2006yi}
P.~Svrcek and E.~Witten,
JHEP \textbf{06}, 051 (2006)
[arXiv:hep-th/0605206 [hep-th]].

\bibitem{Arvanitaki:2009fg}
A.~Arvanitaki, S.~Dimopoulos, S.~Dubovsky, N.~Kaloper and J.~March-Russell,
Phys. Rev. D \textbf{81}, 123530 (2010)
[arXiv:0905.4720 [hep-th]].


\bibitem{Marsh:2015xka}
D.~J.~E.~Marsh,
Phys. Rept. \textbf{643}, 1-79 (2016)
[arXiv:1510.07633 [astro-ph.CO]].








\bibitem{Sikivie:1983ip}
P.~Sikivie,
Phys. Rev. Lett. \textbf{51}, 1415-1417 (1983)
[erratum: Phys. Rev. Lett. \textbf{52}, 695 (1984)].

\bibitem{Raffelt:1987im}
G.~Raffelt and L.~Stodolsky,
Phys. Rev. D \textbf{37}, 1237 (1988).


\bibitem{Carroll:1989vb}
S.~M.~Carroll, G.~B.~Field and R.~Jackiw,
Phys. Rev. D \textbf{41}, 1231 (1990).



\bibitem{Harari:1992ea}
D.~Harari and P.~Sikivie,
Phys. Lett. B \textbf{289}, 67-72 (1992).


\bibitem{Carroll:1998zi}
S.~M.~Carroll,
Phys. Rev. Lett. \textbf{81}, 3067-3070 (1998)
[arXiv:astro-ph/9806099 [astro-ph]].



\bibitem{DeRocco:2018jwe}
W.~DeRocco and A.~Hook,
Phys. Rev. D \textbf{98}, no.3, 035021 (2018)
[arXiv:1802.07273 [hep-ph]].

\bibitem{Obata:2018vvr}
I.~Obata, T.~Fujita and Y.~Michimura,
Phys. Rev. Lett. \textbf{121}, no.16, 161301 (2018)
[arXiv:1805.11753 [astro-ph.CO]].

\bibitem{Liu:2018icu}
H.~Liu, B.~D.~Elwood, M.~Evans and J.~Thaler,
Phys. Rev. D \textbf{100}, no.2, 023548 (2019)
[arXiv:1809.01656 [hep-ph]].


\bibitem{Nagano:2019rbw}
K.~Nagano, T.~Fujita, Y.~Michimura and I.~Obata,
Phys. Rev. Lett. \textbf{123}, no.11, 111301 (2019)
[arXiv:1903.02017 [hep-ph]].



\bibitem{Martynov:2019azm}
D.~Martynov and H.~Miao,
Phys. Rev. D \textbf{101}, no.9, 095034 (2020)
[arXiv:1911.00429 [physics.ins-det]].

\bibitem{Nagano:2021kwx}
K.~Nagano, H.~Nakatsuka, S.~Morisaki, T.~Fujita, Y.~Michimura and I.~Obata,
[arXiv:2106.06800 [hep-ph]].


\bibitem{Michimura:2019qxr}
Y.~Michimura, Y.~Oshima, T.~Watanabe, T.~Kawasaki, H.~Takeda, M.~Ando, K.~Nagano, I.~Obata and T.~Fujita,
J. Phys. Conf. Ser. \textbf{1468}, no.1, 012032 (2020)
[arXiv:1911.05196 [physics.ins-det]].

\bibitem{Oshima:2021irp}
Y.~Oshima, H.~Fujimoto, M.~Ando, T.~Fujita, Y.~Michimura, K.~Nagano, I.~Obata and T.~Watanabe,
[arXiv:2105.06252 [hep-ph]].

\bibitem{Fujimoto:2021uod}
H.~Fujimoto, Y.~Oshima, M.~Ando, T.~Fujita, Y.~Michimura, K.~Nagano and I.~Obata,
[arXiv:2105.08347 [physics.ins-det]].


\bibitem{Fujita:2018zaj}
T.~Fujita, R.~Tazaki and K.~Toma,
Phys. Rev. Lett. \textbf{122}, no.19, 191101 (2019)
[arXiv:1811.03525 [astro-ph.CO]].


\bibitem{Chen:2019fsq}
Y.~Chen, J.~Shu, X.~Xue, Q.~Yuan and Y.~Zhao,
Phys. Rev. Lett. \textbf{124}, no.6, 061102 (2020)
[arXiv:1905.02213 [hep-ph]].

\bibitem{Chen:2021lvo}
Y.~Chen, Y.~Liu, R.~S.~Lu, Y.~Mizuno, J.~Shu, X.~Xue, Q.~Yuan and Y.~Zhao,
[arXiv:2105.04572 [hep-ph]].


\bibitem{Finelli:2008jv}
F.~Finelli and M.~Galaverni,
Phys. Rev. D \textbf{79}, 063002 (2009)
[arXiv:0802.4210 [astro-ph]].

\bibitem{Lee:2013mqa}
S.~Lee, G.~C.~Liu and K.~W.~Ng,
Phys. Rev. D \textbf{89}, no.6, 063010 (2014)
[arXiv:1307.6298 [astro-ph.CO]].

\bibitem{Sigl:2018fba}
G.~Sigl and P.~Trivedi,
[arXiv:1811.07873 [astro-ph.CO]].

\bibitem{Fedderke:2019ajk}
M.~A.~Fedderke, P.~W.~Graham and S.~Rajendran,
Phys. Rev. D \textbf{100}, no.1, 015040 (2019)
[arXiv:1903.02666 [astro-ph.CO]].


\bibitem{Lue:1998mq}
A.~Lue, L.~M.~Wang and M.~Kamionkowski,
Phys. Rev. Lett. \textbf{83}, 1506-1509 (1999)
[arXiv:astro-ph/9812088 [astro-ph]].

\bibitem{Feng:2006dp}
B.~Feng, M.~Li, J.~Q.~Xia, X.~Chen and X.~Zhang,
Phys. Rev. Lett. \textbf{96}, 221302 (2006)
[arXiv:astro-ph/0601095 [astro-ph]].

\bibitem{Liu:2006uh}
G.~C.~Liu, S.~Lee and K.~W.~Ng,
Phys. Rev. Lett. \textbf{97}, 161303 (2006)
[arXiv:astro-ph/0606248 [astro-ph]].

\bibitem{Lee:2014rpa}
S.~Lee, G.~C.~Liu and K.~W.~Ng,
Phys. Lett. B \textbf{746}, 406-409 (2015)
[arXiv:1403.5585 [astro-ph.CO]].

\bibitem{Liu:2016dcg}
G.~C.~Liu and K.~W.~Ng,
Phys. Dark Univ. \textbf{16}, 22-25 (2017)
[arXiv:1612.02104 [astro-ph.CO]].

\bibitem{Lee:2016jym}
S.~Lee, G.~C.~Liu and K.~W.~Ng,
The Universe \textbf{4}, no.4, 29-44 (2016)
[arXiv:1912.12903 [astro-ph.CO]].





\bibitem{Minami:2020odp}
Y.~Minami and E.~Komatsu,
Phys. Rev. Lett. \textbf{125}, no.22, 221301 (2020)
[arXiv:2011.11254 [astro-ph.CO]].




\bibitem{Fujita:2020aqt}
T.~Fujita, Y.~Minami, K.~Murai and H.~Nakatsuka,
Phys. Rev. D \textbf{103}, no.6, 063508 (2021)
[arXiv:2008.02473 [astro-ph.CO]].

\bibitem{Fujita:2020ecn}
T.~Fujita, K.~Murai, H.~Nakatsuka and S.~Tsujikawa,
Phys. Rev. D \textbf{103}, no.4, 043509 (2021)
[arXiv:2011.11894 [astro-ph.CO]].


\bibitem{Berghaus:2020ekh}
K.~V.~Berghaus, P.~W.~Graham, D.~E.~Kaplan, G.~D.~Moore and S.~Rajendran,
[arXiv:2012.10549 [hep-ph]].



\bibitem{Takahashi:2020tqv}
F.~Takahashi and W.~Yin,
JCAP \textbf{04}, 007 (2021)
[arXiv:2012.11576 [hep-ph]].


\bibitem{Sangal:2021qeg}
M.~Sangal, C.~H.~Keitel and M.~Tamburini,
[arXiv:2101.02671 [hep-ph]].


\bibitem{Neves:2021tbt}
M.~J.~Neves, J.~B.~de Oliveira, L.~P.~R.~Ospedal and J.~A.~Helay\"el-Neto,
Phys. Rev. D \textbf{104}, no.1, 015006 (2021)
[arXiv:2101.03642 [hep-th]].


\bibitem{Nagata:2021pvc}
R.~Nagata and T.~Namikawa,
PTEP \textbf{2021}, 053
[arXiv:2102.00133 [astro-ph.CO]].


\bibitem{Fung:2021wbz}
L.~W.~Fung, L.~Li, T.~Liu, H.~N.~Luu, Y.~C.~Qiu and S.~H.~H.~Tye,
[arXiv:2102.11257 [hep-ph]].


\bibitem{Mehta:2021pwf}
V.~M.~Mehta, M.~Demirtas, C.~Long, D.~J.~E.~Marsh, L.~McAllister and M.~J.~Stott,
[arXiv:2103.06812 [hep-th]].



\bibitem{Nakagawa:2021nme}
S.~Nakagawa, F.~Takahashi and M.~Yamada,
[arXiv:2103.08153 [hep-ph]].


\bibitem{Jain:2021shf}
M.~Jain, A.~J.~Long and M.~A.~Amin,
[arXiv:2103.10962 [astro-ph.CO]].


\bibitem{Tsujikawa:2021rpg}
S.~Tsujikawa,
Phys. Rev. D \textbf{103}, no.12, 123533 (2021)
[arXiv:2103.12342 [astro-ph.CO]].


\bibitem{Lague:2021frh}
A.~Lagu\"e, J.~R.~Bond, R.~Hlo\v{z}ek, K.~K.~Rogers, D.~J.~E.~Marsh and D.~Grin,
[arXiv:2104.07802 [astro-ph.CO]].


\bibitem{Reig:2021ipa}
M.~Reig,
[arXiv:2104.09923 [hep-ph]].



\bibitem{Clark:2021kze}
S.~E.~Clark, C.~G.~Kim, J.~C.~Hill and B.~S.~Hensley,
[arXiv:2105.00120 [astro-ph.GA]].


\bibitem{Fung:2021fcj}
L.~W.~Fung, L.~Li, T.~Liu, H.~N.~Luu, Y.~C.~Qiu and S.~H.~H.~Tye,
[arXiv:2105.01631 [astro-ph.CO]].


\bibitem{Namikawa:2021gbr}
T.~Namikawa,
[arXiv:2105.03367 [astro-ph.CO]].


\bibitem{Kim:2021eye}
D.~Kim, Y.~Kim, Y.~K.~Semertzidis, Y.~C.~Shin and W.~Yin,
[arXiv:2105.03422 [hep-ph]].


\bibitem{Perivolaropoulos:2021jda}
L.~Perivolaropoulos and F.~Skara,
[arXiv:2105.05208 [astro-ph.CO]].

\bibitem{Alvey:2021hjp}
J.~Alvey and M.~Escudero Abenza,
[arXiv:2106.04226 [hep-th]].


\bibitem{Choi:2021aze}
G.~Choi, W.~Lin, L.~Visinelli and T.~T.~Yanagida,
[arXiv:2106.12602 [hep-ph]].


\bibitem{Berera:2021xqa}
A.~Berera, S.~Brahma, R.~Brandenberger, J.~Calder\'on-Figueroa and A.~Heavens,
[arXiv:2107.06914 [hep-ph]].


\bibitem{Fukugita:1994hq}
M.~Fukugita and T.~Yanagida,
YITP-K-1098.


\bibitem{Frieman:1995pm}
J.~A.~Frieman, C.~T.~Hill, A.~Stebbins and I.~Waga,
Phys. Rev. Lett. \textbf{75}, 2077-2080 (1995)
[arXiv:astro-ph/9505060 [astro-ph]].


\bibitem{Kim:1998kx}
J.~E.~Kim,
JHEP \textbf{05}, 022 (1999)
[arXiv:hep-ph/9811509 [hep-ph]].

\bibitem{Kim:1999dc}
J.~E.~Kim,
JHEP \textbf{06}, 016 (2000)
[arXiv:hep-ph/9907528 [hep-ph]].


\bibitem{Choi:1999xn}
K.~Choi,
Phys. Rev. D \textbf{62}, 043509 (2000)
[arXiv:hep-ph/9902292 [hep-ph]].



\bibitem{Nomura:2000yk}
Y.~Nomura, T.~Watari and T.~Yanagida,
Phys. Lett. B \textbf{484}, 103-111 (2000)
[arXiv:hep-ph/0004182 [hep-ph]].



\bibitem{Kim:2002tq}
J.~E.~Kim and H.~P.~Nilles,
Phys. Lett. B \textbf{553}, 1-6 (2003)
[arXiv:hep-ph/0210402 [hep-ph]].


\bibitem{Copeland:2006wr}
E.~J.~Copeland, M.~Sami and S.~Tsujikawa,
Int. J. Mod. Phys. D \textbf{15}, 1753-1936 (2006)
[arXiv:hep-th/0603057 [hep-th]].



\bibitem{Banks:2003sx}
T.~Banks, M.~Dine, P.~J.~Fox and E.~Gorbatov,
JCAP \textbf{06}, 001 (2003)
[arXiv:hep-th/0303252 [hep-th]].



\bibitem{Kaloper:2005aj}
N.~Kaloper and L.~Sorbo,
JCAP \textbf{04}, 007 (2006)
[arXiv:astro-ph/0511543 [astro-ph]].


\bibitem{Kim:2009cp}
J.~E.~Kim and H.~P.~Nilles,
JCAP \textbf{05}, 010 (2009)
[arXiv:0902.3610 [hep-th]].



\bibitem{Chatzistavrakidis:2012bb}
A.~Chatzistavrakidis, E.~Erfani, H.~P.~Nilles and I.~Zavala,
JCAP \textbf{09}, 006 (2012)
[arXiv:1207.1128 [hep-ph]].






\bibitem{Kim:2004rp}
J.~E.~Kim, H.~P.~Nilles and M.~Peloso,
JCAP \textbf{01}, 005 (2005)
[arXiv:hep-ph/0409138 [hep-ph]].


\bibitem{Dimopoulos:2005ac}
S.~Dimopoulos, S.~Kachru, J.~McGreevy and J.~G.~Wacker,
JCAP \textbf{08}, 003 (2008)
[arXiv:hep-th/0507205 [hep-th]].


\bibitem{Cicoli:2014sva}
M.~Cicoli, K.~Dutta and A.~Maharana,
JCAP \textbf{08}, 012 (2014)
[arXiv:1401.2579 [hep-th]].


\bibitem{Czerny:2014wza}
M.~Czerny and F.~Takahashi,
Phys. Lett. B \textbf{733}, 241-246 (2014)
[arXiv:1401.5212 [hep-ph]].


\bibitem{Choi:2014rja}
K.~Choi, H.~Kim and S.~Yun,
Phys. Rev. D \textbf{90}, 023545 (2014)
[arXiv:1404.6209 [hep-th]].


\bibitem{Kappl:2014lra}
R.~Kappl, S.~Krippendorf and H.~P.~Nilles,
Phys. Lett. B \textbf{737}, 124-128 (2014)
[arXiv:1404.7127 [hep-th]].


\bibitem{Long:2014dta}
C.~Long, L.~McAllister and P.~McGuirk,
Phys. Rev. D \textbf{90}, 023501 (2014)
[arXiv:1404.7852 [hep-th]].




\bibitem{Planck:2018vyg}
N.~Aghanim \textit{et al.} [Planck],
Astron. Astrophys. \textbf{641}, A6 (2020)
[arXiv:1807.06209 [astro-ph.CO]].


\bibitem{Caldwell:2005tm}
R.~R.~Caldwell and E.~V.~Linder,
Phys. Rev. Lett. \textbf{95}, 141301 (2005)
[arXiv:astro-ph/0505494 [astro-ph]].

\bibitem{Scherrer:2007pu}
R.~J.~Scherrer and A.~A.~Sen,
Phys. Rev. D \textbf{77}, 083515 (2008)
[arXiv:0712.3450 [astro-ph]].


\bibitem{Chevallier:2000qy}
M.~Chevallier and D.~Polarski,
Int. J. Mod. Phys. D \textbf{10}, 213-224 (2001)
[arXiv:gr-qc/0009008 [gr-qc]].

\bibitem{Linder:2002et}
E.~V.~Linder,
Phys. Rev. Lett. \textbf{90}, 091301 (2003)
[arXiv:astro-ph/0208512 [astro-ph]].


\bibitem{Hui:2016ltb}
L.~Hui, J.~P.~Ostriker, S.~Tremaine and E.~Witten,
Phys. Rev. D \textbf{95}, no.4, 043541 (2017)
[arXiv:1610.08297 [astro-ph.CO]].


\bibitem{Preskill:1982cy}
J.~Preskill, M.~B.~Wise and F.~Wilczek,
Phys. Lett. B \textbf{120}, 127-132 (1983).

\bibitem{Abbott:1982af}
L.~F.~Abbott and P.~Sikivie,
Phys. Lett. B \textbf{120}, 133-136 (1983).

\bibitem{Dine:1982ah}
M.~Dine and W.~Fischler,
Phys. Lett. B \textbf{120}, 137-141 (1983).

\bibitem{Arias:2012az}
P.~Arias, D.~Cadamuro, M.~Goodsell, J.~Jaeckel, J.~Redondo and A.~Ringwald,
JCAP \textbf{06}, 013 (2012)
[arXiv:1201.5902 [hep-ph]].



\bibitem{Marsh:2010wq}
D.~J.~E.~Marsh and P.~G.~Ferreira,
Phys. Rev. D \textbf{82}, 103528 (2010)
[arXiv:1009.3501 [hep-ph]].


\bibitem{Ferreira:2020fam}
E.~G.~M.~Ferreira,
``Ultra-Light Dark Matter,''
[arXiv:2005.03254 [astro-ph.CO]].


\bibitem{Zioutas:2004hi} 
  K.~Zioutas {\it et al.} [CAST Collaboration],
  Phys.\ Rev.\ Lett.\  {\bf 94}, 121301 (2005)
  [hep-ex/0411033].

\bibitem{Anastassopoulos:2017ftl} 
  V.~Anastassopoulos {\it et al.} [CAST Collaboration],
  Nature Phys.\  {\bf 13}, 584 (2017)
  [arXiv:1705.02290 [hep-ex]].



\bibitem{ADMX:2009iij}
S.~J.~Asztalos \textit{et al.} [ADMX],
Phys. Rev. Lett. \textbf{104}, 041301 (2010)
[arXiv:0910.5914 [astro-ph.CO]].


\bibitem{ADMX:2018gho}
N.~Du \textit{et al.} [ADMX],
Phys. Rev. Lett. \textbf{120}, no.15, 151301 (2018)
[arXiv:1804.05750 [hep-ex]].


\bibitem{ADMX:2020hay}
C.~Bartram \textit{et al.} [ADMX],
Phys. Rev. D \textbf{103}, no.3, 032002 (2021)
[arXiv:2010.06183 [astro-ph.CO]].


\bibitem{Gramolin:2020ict}
A.~V.~Gramolin, D.~Aybas, D.~Johnson, J.~Adam and A.~O.~Sushkov,
Nature Phys. \textbf{17}, no.1, 79-84 (2021)
[arXiv:2003.03348 [hep-ex]].


\bibitem{Salemi:2021gck}
C.~P.~Salemi, J.~W.~Foster, J.~L.~Ouellet, A.~Gavin, K.~M.~W.~Pappas, S.~Cheng, K.~A.~Richardson, R.~Henning, Y.~Kahn and R.~Nguyen, \textit{et al.}
[arXiv:2102.06722 [hep-ex]].


\bibitem{Payez:2014xsa}
A.~Payez, C.~Evoli, T.~Fischer, M.~Giannotti, A.~Mirizzi and A.~Ringwald,
JCAP \textbf{02}, 006 (2015)
[arXiv:1410.3747 [astro-ph.HE]].




\bibitem{Reynolds:2019uqt}
C.~S.~Reynolds, M.~C.~D.~Marsh, H.~R.~Russell, A.~C.~Fabian, R.~Smith, F.~Tombesi and S.~Veilleux,
[arXiv:1907.05475 [hep-ph]].


\bibitem{Vogel:2013bta} 
  J.~K.~Vogel {\it et al.},
  arXiv:1302.3273 [physics.ins-det].


\bibitem{Irastorza:2018dyq}
I.~G.~Irastorza and J.~Redondo,
Prog. Part. Nucl. Phys. \textbf{102}, 89-159 (2018)
[arXiv:1801.08127 [hep-ph]].


\bibitem{Bahre:2013ywa}
R.~B\"ahre, B.~D\"obrich, J.~Dreyling-Eschweiler, S.~Ghazaryan, R.~Hodajerdi, D.~Horns, F.~Januschek, E.~A.~Knabbe, A.~Lindner and D.~Notz, \textit{et al.}
JINST \textbf{8}, T09001 (2013)
[arXiv:1302.5647 [physics.ins-det]].



\bibitem{Kahn:2016aff} 
  Y.~Kahn, B.~R.~Safdi and J.~Thaler,
  Phys.\ Rev.\ Lett.\  {\bf 117}, no. 14, 141801 (2016)
  [arXiv:1602.01086 [hep-ph]].



\bibitem{Conlon:2017ofb}
J.~P.~Conlon, F.~Day, N.~Jennings, S.~Krippendorf and F.~Muia,
Mon. Not. Roy. Astron. Soc. \textbf{473}, no.4, 4932-4936 (2018)
[arXiv:1707.00176 [astro-ph.HE]].


\bibitem{Gross:1980br}
D.~J.~Gross, R.~D.~Pisarski and L.~G.~Yaffe,
Rev. Mod. Phys. \textbf{53}, 43 (1981).

\bibitem{Wantz:2009it}
O.~Wantz and E.~P.~S.~Shellard,
Phys. Rev. D \textbf{82}, 123508 (2010)
[arXiv:0910.1066 [astro-ph.CO]].


\bibitem{Planck:2018jri}
Y.~Akrami \textit{et al.} [Planck],
Astron. Astrophys. \textbf{641}, A10 (2020)
[arXiv:1807.06211 [astro-ph.CO]].


\bibitem{Agrawal:2018mkd}
P.~Agrawal, J.~Fan and M.~Reece,
JHEP \textbf{10}, 193 (2018)
[arXiv:1806.09621 [hep-th]].






\bibitem{Babu:1994id}
K.~S.~Babu, S.~M.~Barr and D.~Seckel,
Phys. Lett. B \textbf{336}, 213-220 (1994)
[arXiv:hep-ph/9406308 [hep-ph]].

\bibitem{Bachlechner:2014hsa}
T.~C.~Bachlechner, M.~Dias, J.~Frazer and L.~McAllister,
Phys. Rev. D \textbf{91}, no.2, 023520 (2015)
[arXiv:1404.7496 [hep-th]].

\bibitem{Shiu:2015uva}
G.~Shiu, W.~Staessens and F.~Ye,
Phys. Rev. Lett. \textbf{115}, 181601 (2015)
[arXiv:1503.01015 [hep-th]].




















\bibitem{Blumenhagen:2018hsh}
R.~Blumenhagen,
PoS \textbf{CORFU2017}, 175 (2018)
[arXiv:1804.10504 [hep-th]].

\bibitem{Cicoli:2018kdo}
M.~Cicoli, S.~De Alwis, A.~Maharana, F.~Muia and F.~Quevedo,
Fortsch. Phys. \textbf{67}, no.1-2, 1800079 (2019)
[arXiv:1808.08967 [hep-th]].


\bibitem{Ibe:2018ffn}
M.~Ibe, M.~Yamazaki and T.~T.~Yanagida,
Class. Quant. Grav. \textbf{36}, no.23, 235020 (2019)
[arXiv:1811.04664 [hep-th]].



\bibitem{Palti:2019pca}
E.~Palti,
Fortsch. Phys. \textbf{67}, no.6, 1900037 (2019)
[arXiv:1903.06239 [hep-th]].


\bibitem{Banerjee:2020xcn}
A.~Banerjee, H.~Cai, L.~Heisenberg, E.~\'O.~Colg\'ain, M.~M.~Sheikh-Jabbari and T.~Yang,
Phys. Rev. D \textbf{103}, no.8, L081305 (2021)
[arXiv:2006.00244 [astro-ph.CO]].







\end{thebibliography}
\end{document}